\documentclass[%
 aip,
 amsmath,amssymb,
 reprint,%
]{revtex4-1}

\usepackage{graphicx}
\usepackage{dcolumn}
\usepackage{bm}

\usepackage[utf8]{inputenc}
\usepackage[T1]{fontenc}
\usepackage{mathptmx}

\usepackage{color,soul} 

\begin{document}

\preprint{AIP/123-QED}

\title{Theoretical Insights into Mechanisms of Stochastic Gating in Channel-Facilitated Molecular Transport}

\author{Aram Davtyan}
 \affiliation{Center for Theoretical Biological Physics, Rice University, Houston, Texas 77005, USA}
\author{Anatoly B. Kolomeisky}%
 \email{tolya@rice.edu}
 \affiliation{Center for Theoretical Biological Physics, Rice University, Houston, Texas 77005, USA}
 \affiliation{Department of Chemistry, Rice University, Houston, Texas 77005-1892, USA}
 \affiliation{Department of Chemical and Biomolecular Engineering, Rice University, Houston, Texas 77005-1892, USA}%

\date{\today}

\begin{abstract}

Molecular motion through pores plays a crucial role in various natural and industrial processes. One of the most fascinating features of biological channel-facilitated transport is a stochastic gating process, when the channels dynamically fluctuate between several conformations during the translocation. Although this phenomenon has been intensively investigated, many properties of translocation in dynamically changing environment remain not well understood microscopically. We developed a discrete-state stochastic framework to analyze the molecular mechanisms of transport processes with stochastic gating by explicitly calculating molecular fluxes through the pores. Two scenarios are specifically investigated: 1) symmetry preserving stochastic gating with free-energy changes, and 2) stochastic gating with symmetry changes but without modifications in the overall particle-pore interactions. It is found that stochastic gating can both accelerate or slow down the molecular translocation depending on the specific parameters of the system. We argue that biological systems might optimize their performance by utilizing conformational fluctuations of channels. Our theoretical analysis clarifies physical-chemical aspects of the molecular mechanisms of transport with stochastic gating.

\end{abstract}

\maketitle


\section{\label{sec:intro} Introduction}

Molecular transport via channels is critically important in multiple biological processes where metabolites and nutrients must be moved between different cellular compartments and delivered to specific locations.\cite{lodish2008molecular} It is also crucial in many industrial processes, e.g., in those that involve the separation of chemical mixtures and water purification.\cite{meloan1999chemical,sparreboom2009principles} The importance of translocation through pores stimulated extensive theoretical studies to uncover the underlying molecular mechanisms.\cite{Berezhkovskii2002,Berezhkovskii2005,Kolomeisky2007,Bhatia2011,Kolomeisky2011,Maffeo2012,Agah2015,muthukumar2016polymer,roux2004theoretical,sparreboom2009principles} But many questions remain open. Specifically, most of existing theoretical studies of channel-facilitated molecular transport  concentrate on investigating systems, where interactions between particles and the channel are constant over the time. However, biological cells are very dynamic non-equilibrium systems, where intermolecular interactions frequently change as a result of passive or active regulation processes. For instance, ion channels are largely regulated in biological cells by varying the membrane potentials and by changing the dynamics of ligands binding to membrane receptors.\cite{hille2001ion} As a result, the channel can undergo significant  conformational changes that might close or restrict the passage of particles through it for some periods of time. This is known as a {\it stochastic gating} phenomenon, and it is widely observed in biological systems.\cite{lodish2008molecular}

Because it is extremely difficult to account for all processes in channel transport at the atomistic level, most theoretical investigations follow coarse-grained, mesoscopic approaches, which can be divided into two main categories.\cite{Kolomeisky2007,Kolomeisky2011,Berezhkovskii2002,Berezhkovskii2005,sparreboom2009principles} In one of them, the channel transport is studied using a continuum diffusion model. It views the translocation as a quasi one-dimensional motion in the effective potential created by interactions between the molecules and the pores.\cite{Berezhkovskii2002,Berezhkovskii2005} In cases when these effective potentials (and thus their effect on particle diffusion) can be reasonably well evaluated, a quantitative description of the molecular translocation through pores can be obtained using this methodology. An alternative approach employs a discrete chemical-kinetic description, where the molecular transport is represented as a sequence of chemical transitions between different states that correspond to minima in the interaction potential (free-energy) profile.\cite{Kolomeisky2007,Kolomeisky2011,Agah2015} The advantage of this approach is that some of these transition rates can be measured in experiments on channel transport. A comprehensive theoretical framework for investigation of chemical mechanisms of translocation and selectivity under stationary-state conditions was recently developed based on this discrete-state kinetic approach.\cite{Kolomeisky2007,Kolomeisky2011,Agah2015} Importantly, it was also shown that both theoretical methods are mathematically equivalent.\cite{Kolomeisky2007,Kolomeisky2011}

Recently, stochastic gating has been investigated theoretically using the continuum diffusion description.\cite{Berezhkovskii2017,Berezhkovskii2018} It was shown that the stochastic gating can be successfully used as a selectivity mechanisms for molecular translocation through pores, and the dynamics of gating might strongly influence the channel transport. As a complementary approach, in this work we developed a simple theory of stochastic gating for particles traveling through molecular channels using the discrete-state chemical-kinetic approach. Our goal is to understand the general features of the stochastic gating and how it can optimize the molecular transport. For this reason, we specifically consider two limiting situations: 1) when the stochastic gating is associated with fluctuations in the free-energy for a pore system that is always symmetric; and 2) when the stochastic gating changes the symmetry of the interaction potential without  overall modifications in the interaction strength between the molecule and the pore. In both cases, we are asking the following questions. Are changes associated with the stochastic gating beneficial for the transport through the channel, i.e., do they increase the flux relative to the stationary system without stochastic gating? Are there optimal conditions, such as particle concentrations outside the channel, system transition rates, and particle-channel interaction energies, that might maximize or minimize the flux? Is there a possibility for a stochastic resonance, i.e., is there a special rate of conformational transitions that leads to the maximal particle current?

The paper is organized as follows. In Section \ref{sec:res}, we specifically analyze two different stochastic gating models. Using a simple chemical-kinetic theory, we analytically solve for stationary properties of the systems and determine the particle fluxes through the channel. Analytical results are utilized then to deduce the molecular features of the system and the role of stochastic gating in the channel transport. Section ~\ref{sec:conc} provides summary and concluding remarks.

\section{\label{sec:res} Theoretical Method and Results}


\subsection{Stochastic Gating with Free-Energy Change}

\begin{figure*}
  \centering
    \includegraphics[width=0.9\textwidth]{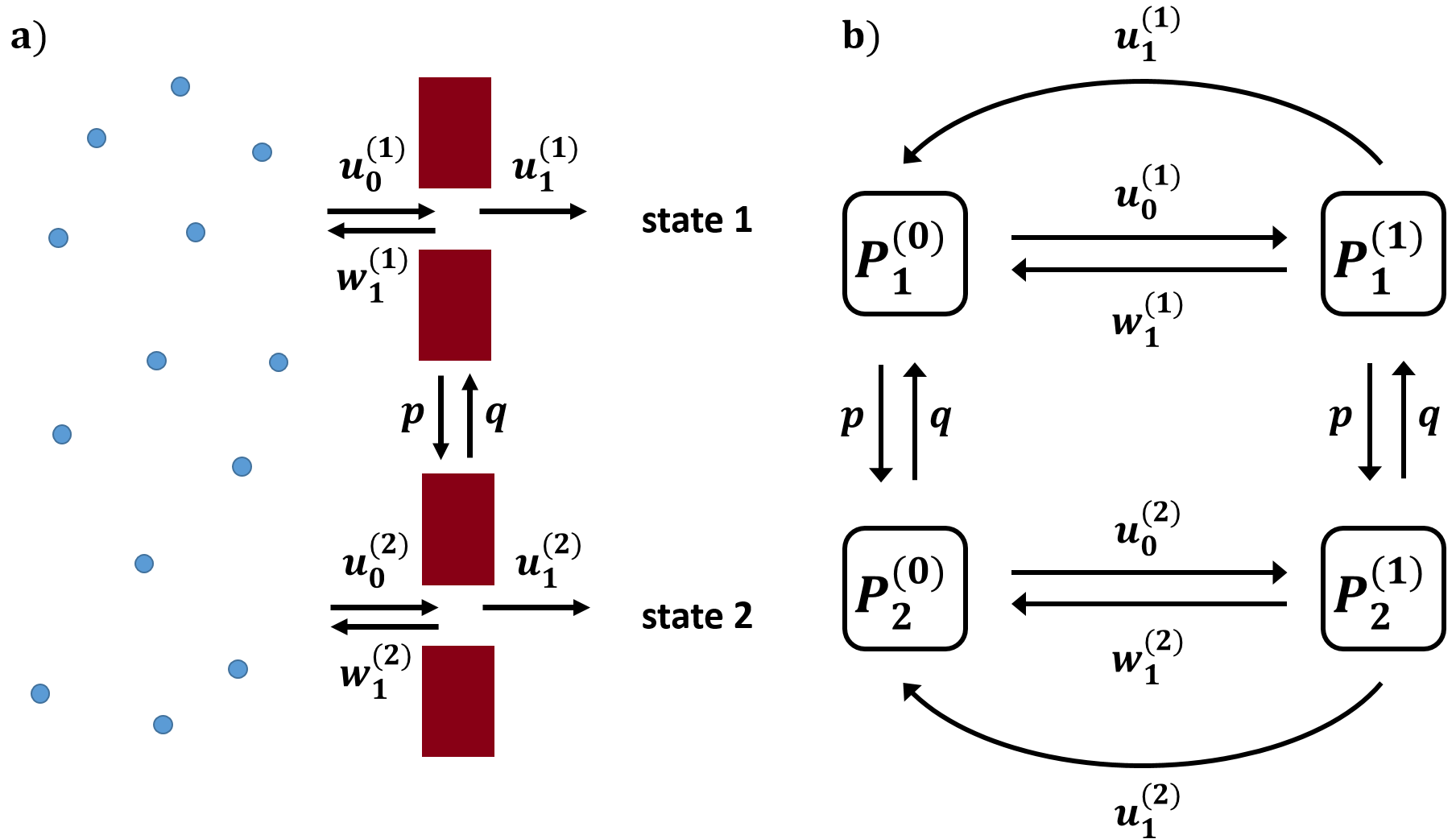}
  \caption{\label{fig:prob1} a) A schematic view of the molecular transport via a conformationally fluctuating channel in the model of stochastic gating with free-energy changes. b) A corresponding chemical-kinetic diagram for the model.}
\end{figure*}

Let us consider a molecular translocation via a pore as shown in Figure~\ref{fig:prob1}. It is assumed that there is a constant concentration gradient between two sides of the channel, i.e., the concentration of molecules to the left is taken to be equal to $c$ at all times, while the concentration on the right is always equal to zero. In addition, only a single molecule can be found inside the pore, or the channel can be empty. This corresponds to very strong repulsions between the particles.\cite{Berezhkovskii2017} We also assume that the channel interacts with the translocating molecule, and the pore can stochastically switch between two conformational states (labeled as state 1 or state 2), where this interaction differs: see Figure~\ref{fig:prob1}. We denote the forward and backward transition rates between the states 1 and 2 as $p$ and $q$, respectively (Figure~\ref{fig:prob1}). The difference in interaction energies between the channel and the particle in both conformations is labeled as $E$, and it can take both positive and negative values. The entrance rate to the channel when it is in the state $i$ is equal to $u_0^{(i)}$, and it is proportional to the concentration $c$: $u_0^{(i)}=c k_{on}^{(i)}$, where $i=1,2$. The exit rate of the pore-bound particle to pass to the right of the pore is equal to $u_1^{(i)}$, while the rate to exit back to the left of the channel is given by $w_1^{(i)}$: see Figure~\ref{fig:prob1}.

We define a function $P_i^{(j)}(t)$ as a probability to find the system at time $t$ in a state  in which the channel is in the conformational state $i$ ($i=1$ or 2) and the pore occupation is given by the state $j$ ($j=0$ corresponds to the empty channel, and $j=1$ describes the particle in the channel). The temporal evolution of these probabilities is controlled by the following set of forward master equations:
\begin{equation}
\frac{dP_1^{(0)}(t)}{dt}=-(u_0^{(1)} + p) P_1^{(0)}(t) + (u_1^{(1)} + w_1^{(1)}) P_1^{(1)}(t) + q P_2^{(0)}(t)
\label{eq:mastp10},
\end{equation}
\begin{equation}
\frac{dP_1^{(1)}(t)}{dt}=-(u_1^{(1)} + w_1^{(1)} + p) P_1^{(1)}(t) + u_0^{(1)} P_1^{(0)}(t) + q P_2^{(1)}(t)
\label{eq:mastp11},
\end{equation}
\begin{equation}
\frac{dP_2^{(0)}(t)}{dt}=-(u_0^{(2)} + q) P_2^{(0)}(t) + (u_1^{(2)} + w_1^{(2)}) P_2^{(1)}(t) + p P_1^{(0)}(t)
\label{eq:mastp20},
\end{equation}
\begin{equation}
\frac{dP_2^{(1)}(t)}{dt}=-(u_1^{(2)} + w_1^{(2)} + q) P_2^{(1)}(t) + u_0^{(2)} P_2^{(0)}(t) + p P_1^{(1)}(t)
\label{eq:mastp21}.
\end{equation}
In addition, the normalization requires that at all times we have
\begin{equation}
P_1^{(0)}(t) + P_1^{(1)}(t) + P_2^{(0)}(t) + P_2^{(1)}(t) = 1.
\label{eq:pnorm}
\end{equation}

We are interested in stationary solutions, when $\frac{dP_i^{(j)}(t)}{dt}=0$. For this case, the set of  Equations (\ref{eq:mastp10}-\ref{eq:pnorm}) can be solved analytically (note that only four of these five equations are independent), and the following expressions for stationary probabilities $P_i^{(j)}$ can be obtained:
\begin{small}
\begin{eqnarray}
P_1^{(0)} = \frac{1}{p+q} \times & & \nonumber \\
& \frac{q\left[ (u_1^{(1)} + w_1^{(1)})(q + u_0^{(2)} + u_1^{(2)} + w_1^{(2)}) + p(u_1^{(2)} + w_1^{(2)}) \right]} {\left[ (u_0^{(1)} + u_1^{(1)} + w_1^{(1)}) (q + u_0^{(2)} + u_1^{(2)} + w_1^{(2)}) + p (u_0^{(2)} + u_1^{(2)} + w_1^{(2)}) \right]}  &
\label{eq:solp10},
\end{eqnarray}
\begin{eqnarray}
P_1^{(1)} = \frac{1}{p+q} \times & & \nonumber \\
&\frac{q \left[u_0^{(1)}(q + u_0^{(2)} + u_1^{(2)} + w_1^{(2)}) + p u_0^{(2)}\right]} {\left[ (u_0^{(1)} + u_1^{(1)} + w_1^{(1)}) (q + u_0^{(2)} + u_1^{(2)} + w_1^{(2)}) + p (u_0^{(2)} + u_1^{(2)} + w_1^{(2)})\right]}&
\label{eq:solp11},
\end{eqnarray}
\begin{eqnarray}
P_2^{(0)} = \frac{1}{p+q}\times \nonumber & &\\
& \frac{p\left[ (u_1^{(1)} + w_1^{(1)})(q + u_1^{(2)} + w_1^{(2)}) + (p + u_0^{(1)})(u_1^{(2)} + w_1^{(2)}) \right]} {\left[ (u_0^{(1)} + u_1^{(1)} + w_1^{(1)}) (q + u_0^{(2)} + u_1^{(2)} + w_1^{(2)}) + p (u_0^{(2)} + u_1^{(2)} + w_1^{(2)}) \right]}&
\label{eq:solp20},
\end{eqnarray}
\begin{eqnarray}
P_2^{(1)} = \frac{1}{p+q}\times & &\nonumber \\
& \frac{p\left[ u_0^{(1)}q + (p + u_0^{(1)} + u_1^{(1)} + w_1^{(1)}) u_0^{(2)} \right]} {\left[ (u_0^{(1)} + u_1^{(1)} + w_1^{(1)}) (q + u_0^{(2)} + u_1^{(2)} + w_1^{(2)}) + p (u_0^{(2)} + u_1^{(2)} + w_1^{(2)}) \right]} &
\label{eq:solp21}.
\end{eqnarray}
\end{small}

Given these analytical expressions, we now can evaluate the molecular flux through the channel in terms of the transitions rates via
\begin{equation}
J = u_1^{(1)} P_1^{(1)} + u_1^{(2)} P_2^{(1)}
\label{eq:fluxdef}.
\end{equation}
This gives the total particle flux of leaving the channel into the right chamber from both possible channel conformations. Substituting Equations (\ref{eq:solp11}) and (\ref{eq:solp21}) into Equation (\ref{eq:fluxdef}), we obtain the following general analytical expression for the particle current:
\begin{widetext}
\begin{equation}
J = \frac{p(p + u_0^{(1)} + u_1^{(1)} + w_1^{(1)}) u_0^{(2)} u_1^{(2)} + q(q + u_0^{(2)} + u_1^{(2)} + w_1^{(2)}) u_0^{(1)} u_1^{(1)} + p q (u_0^{(1)} u_1^{(2)} + u_1^{(1)} u_0^{(2)})} {(p+q)\left[ (u_0^{(1)} + u_1^{(1)} + w_1^{(1)}) (q + u_0^{(2)} + u_1^{(2)} + w_1^{(2)}) + p (u_0^{(2)} + u_1^{(2)} + w_1^{(2)}) \right]}
\label{eq:fluxfull}.
\end{equation}
\end{widetext}

The transition rates in the system are not independent, and they are connected to each other via detailed balance-like relations, which can be stated in the following form:
\begin{equation}
\frac{u_0^{(2)}}{w_1^{(2)}} = \frac{u_0^{(1)}}{w_1^{(1)}} e^{\beta E}, \ \ \ \ \  
\frac{u_1^{(2)}}{u_1^{(1)}} =  e^{-\beta E}, \ \ \ \ \ 
\frac{p}{q} =  e^{\beta E}
\label{eq:detbalres}.
\end{equation}
The physical meaning of these equations is simple: in the conformational state 2 the particle has energy lower by $E$ (if $E>0$), and the transitions to the states with lower free energy are faster, while the transitions to the states with higher free energy are slower. Similar arguments can be presented for $E<0$.

Using Equation (\ref{eq:detbalres}), we can simplify our notations and rewrite all rates as
\begin{equation}
u_0^{(1)} \equiv u_0, \ \ \ \ \  w_1^{(1)} \equiv w_1, \ \ \ \ \  u_1^{(1)} \equiv u_1
\label{eq:ratesth1},
\end{equation}
\begin{equation}
u_0^{(2)} = u_0 e^{\beta \theta E}, \ \ \ \ \  w_1^{(2)} = w_1 e^{\beta (\theta-1) E}, \ \ \ \ \  u_1^{(2)} = u_1 e^{-\beta E}
\label{eq:ratesth2},
\end{equation}
\begin{equation}
p = p_0 e^{\beta \theta E}, \ \ \ \ \  q = p_0 e^{\beta (\theta-1) E}
\label{eq:ratesth3}.
\end{equation}
Here we introduced a parameter $0 \leqslant \theta \leqslant 1$, which describes the relative effect of the difference in the interaction energies $E$ for forward and backward transition rates. It is assumed that this coefficient is the same for all transitions, which is generally not correct, but relaxing this condition will not change main physical predictions of our model. For convenience, from now on we take $p_0 \equiv p$. Consequently, Equation (\ref{eq:fluxfull}) can be simplified, leading to
\begin{equation}
J = \frac{u_0 u_1}{1 + e^{\beta E}} \frac{2 p (1 + e^{\beta (\theta + 1) E}) e^{\beta \theta E} + U e^{\beta (\theta+1) E} + W}{p (U + W) e^{\beta \theta E} + U W}
\label{eq:fluxsimp1},
\end{equation}
where
\begin{equation}
U = u_0 + u_1 + w_1
\label{eq:Ufull},
\end{equation}
\begin{equation}
W = u_0 e^{\beta (\theta + 1) E} + u_1 + w_1 e^{\beta \theta E}
\label{eq:Wfull}.
\end{equation}
To simplify Equation (\ref{eq:fluxsimp1}) even further, we assume that $\theta=1/2$ - it can be shown that relaxing this condition does not change the physics of the problem. We also define a dimensionless interaction parameter $x \equiv e^{\beta E/2}$. Then, we obtain a final compact expression:
\begin{equation}
J = \frac{u_0 u_1}{1+x^2}\frac{2 p x (1+ x^3) +  U x^3 + W}{p x (U + W) + U W}
\label{eq:fluxsimp2},
\end{equation}
where $U$ is again given by Equation (\ref{eq:Ufull}), and
\begin{equation}
W = u_0 x^3 + u_1 + w_1 x
\label{eq:Wsimp2}.
\end{equation}

First, let us consider the particle current presented in Equation (\ref{eq:fluxsimp2}) in several limiting situations. When the entrance rate is very large ($u_{0} \gg 1$), exiting from the pore will be a rate-limiting step, and the molecular flux has a very simple expression,
\begin{equation}
J \simeq \frac{2 u_{1}}{1+x^{2}}.
\end{equation}
When the exit rate is very large ($u_{1} \gg 1$), the entrance to the channel is a rate-limiting step, and another simple expression for the current can be obtained,
\begin{equation}
J \simeq \frac{2 u_{0}}{1+x^{2}}.
\end{equation}
In both cases, the molecular flux is independent of the backward transition rate $w_{1}$ because the system does not have a chance for such transitions at these limiting cases.

Equation (\ref{eq:fluxsimp2}) can now be analyzed to understand the general features of the molecular transport via the pores. For $x=1$ (or $E=0$) when there is no free-energy differences between two channel conformations, it reduces to:
\begin{equation}
J_0 = \frac{u_0 u_1}{u_0 + u_1 + w_1}
\label{eq:fluxsimp3}.
\end{equation}
This is also the particle current for the system without the stochastic gating since the conformational fluctuations do not affect the particle-pore interactions. For $x=0$ (or $E\to-\infty$) when the molecular flux in the state 2 is completely blocked, we derive
\begin{equation}
J_{-\infty} = \frac{u_0 u_1}{u_0 + u_1 + w_1} = J_0
\label{eq:fluxsimp4}.
\end{equation}
Again, this coincides with the particle current in the system without the stochastic gating. In this case, there is a strong repulsion between the particle and the channel in the conformation 2, and the particle does not enter the pore in this conformation because the system is mostly in the state 1. For $x \to+\infty $ (or $E\to+\infty$) we have
\begin{equation}
J_{+\infty} \approx \frac{2 u_1}{x^2 } \rightarrow 0
\label{eq:fluxsimp5}.
\end{equation}
This result can be explained in the following way. The interaction between the particle and the channel is much more attractive in the state 2, so the system is mostly in this conformation. But then it cannot pass the channel due to strong attractive interactions that trap the molecule inside the pore, and this leads to zero molecular flux at these conditions.

To quantify the effect of stochastic gating, one might consider a normalized current using Equation (\ref{eq:fluxsimp3}),
\begin{equation}
J_n = \frac{J}{J_0} = \frac{U}{1+x^2}\frac{2 p x (1+ x^3) +  U x^3 + W}{p x (U + W) + U W}
\label{eq:fluxnorm}.
\end{equation}
If the normalized current $J_{n}$ is larger than one, then the stochastic gating increases the channel flux, while for the case of $J_{n}<1$, the effect of stochastic gating is to decrease the particle current via the pore.

To understand how modifying the speed of channel conformations affects the molecular transport, we vary the parameter $p$, which is proportional to the rate of conformational fluctuations. It is explicitly shown in Appendix~A that the derivative of $J$ with respect to $p$ is always positive for positively defined transition rates $u_0$, $u_1$, and $w_1$, and for all values of $x$. Thus, in contrast to some naive expectations of optimal speed of conformational transitions, the particle current will always increase monotonically with increasing the frequency of conformational changes. This is a physically clear result since increasing the rate of conformational changes gives the particles more possibilities to cross the channel without being trapped for significant periods of time in energetically unfavorable states.

It is convenient to consider the limiting cases of very slow and very fast conformational changes. When $p \rightarrow 0$ we obtain
\begin{small}
\begin{equation}
J_n(p \rightarrow 0) = \frac{U x^3 + W}{(1+x^2) W} = \frac{2 u_0 x^3 + u_1 (1+x^3) + w_1 x(1+x^2)}{(1+x^2)(u_0 x^3 + u_1 + w_1 x)}
\label{eq:fluxpzero},
\end{equation}
while for $p \rightarrow + \infty$ the normalized flux is given by
\begin{equation}
J_n(p \rightarrow \infty) = \frac{2 (1+x^3)}{1+x^2} \frac{U}{U+W} = \frac{2 (1+x^3)}{1+x^2} \frac{u_0 + u_1 + w_1}{u_0 (1 + x^3) +2  u_1 + w_1 (1+x)}
\label{eq:fluxpinf}.
\end{equation}
\end{small}
One can also see that $J_n(p \rightarrow 0)$ and $J_n(p \to \infty)$ can both take values smaller and larger than 1, however $J_n(p \rightarrow \infty) \ge J_n(p \rightarrow 0)$, with equality only possible for $x=1$ (the situation without stochastic gating). 

\begin{figure*}
  \centering
    \includegraphics[width=1.0\textwidth]{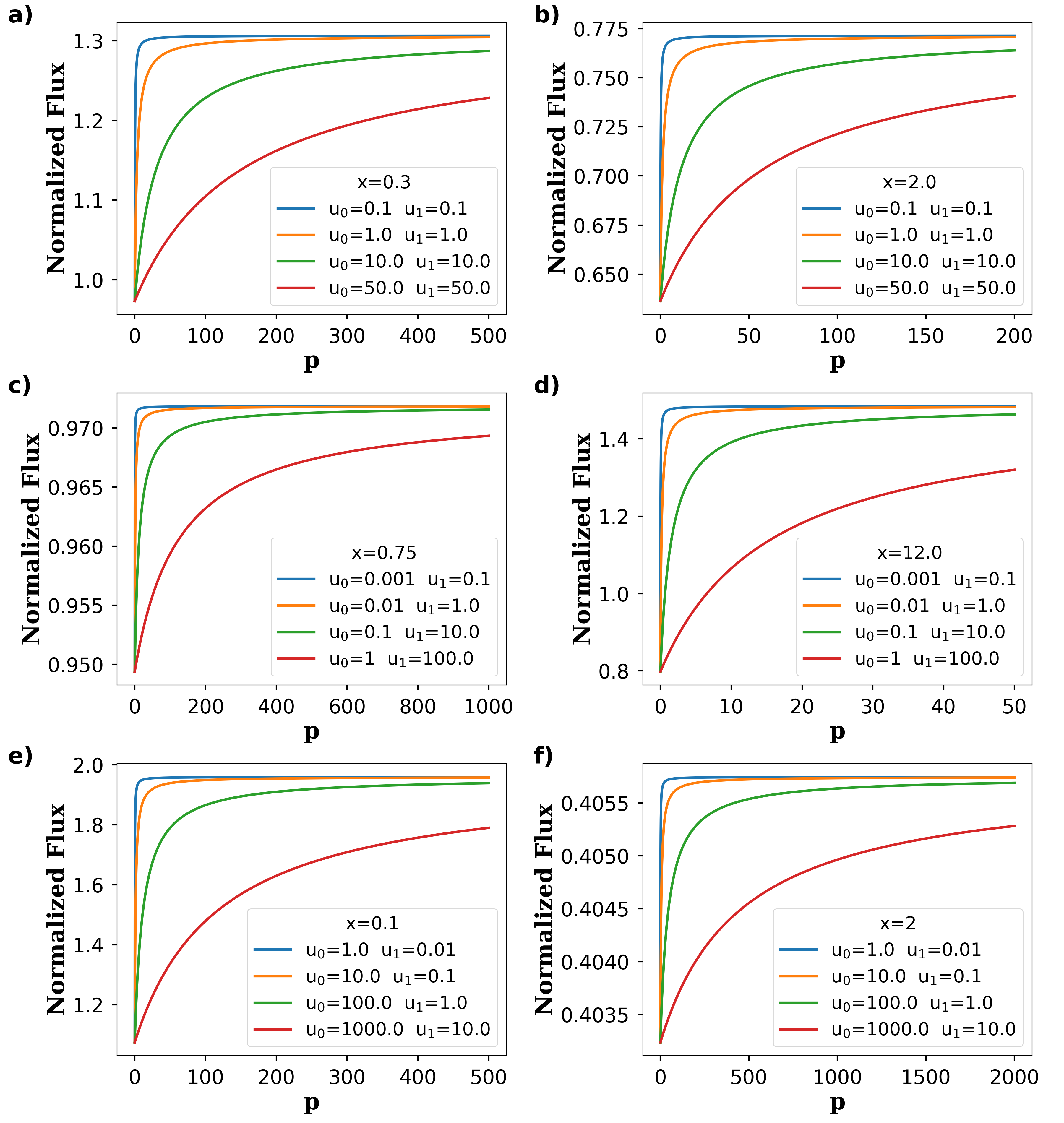}
  \caption{\label{fig:fluxvsp} Plots of normalized current $J/J_0$ as a function of the parameter $p$, which is proportional to the rate of conformational transitions, for the model of the stochastic gating with changing free-energy. For calculations we used: a) and b) $u_0=u_1=w_1$, c) and d) $u_0<<u_1$, $w_1=u_1$, and e) and f) $u_0>>u_1$, $w_1=u_1$.}
\end{figure*}

Our theoretical results for the dependence of the particle current on the speed of conformational fluctuations are presented in Figure~\ref{fig:fluxvsp} for various sets of parameters. As explained above, the molecular fluxes always increase for faster conformational transitions.  In some cases, the normalized current exceeds one (starting below one), suggesting that the stochastic gating can improve the channel-facilitated molecular transport. However, in other cases it is always less than one, and the stochastic gating cannot optimize the molecular fluxes at these conditions. It seems that in most situations the optimization might be achieved for $x<1$ when opening the second energetically less favorable conformation gives molecules another pathway to translocate while not trapping them on their way our of the channel (see Figures 2a and 2e). But there are also ranges of parameters when the stochastic gating might optimize the dynamics for $x>1$ (Figure 2d). 

\begin{figure*}
  \centering
    \includegraphics[width=1.0\textwidth]{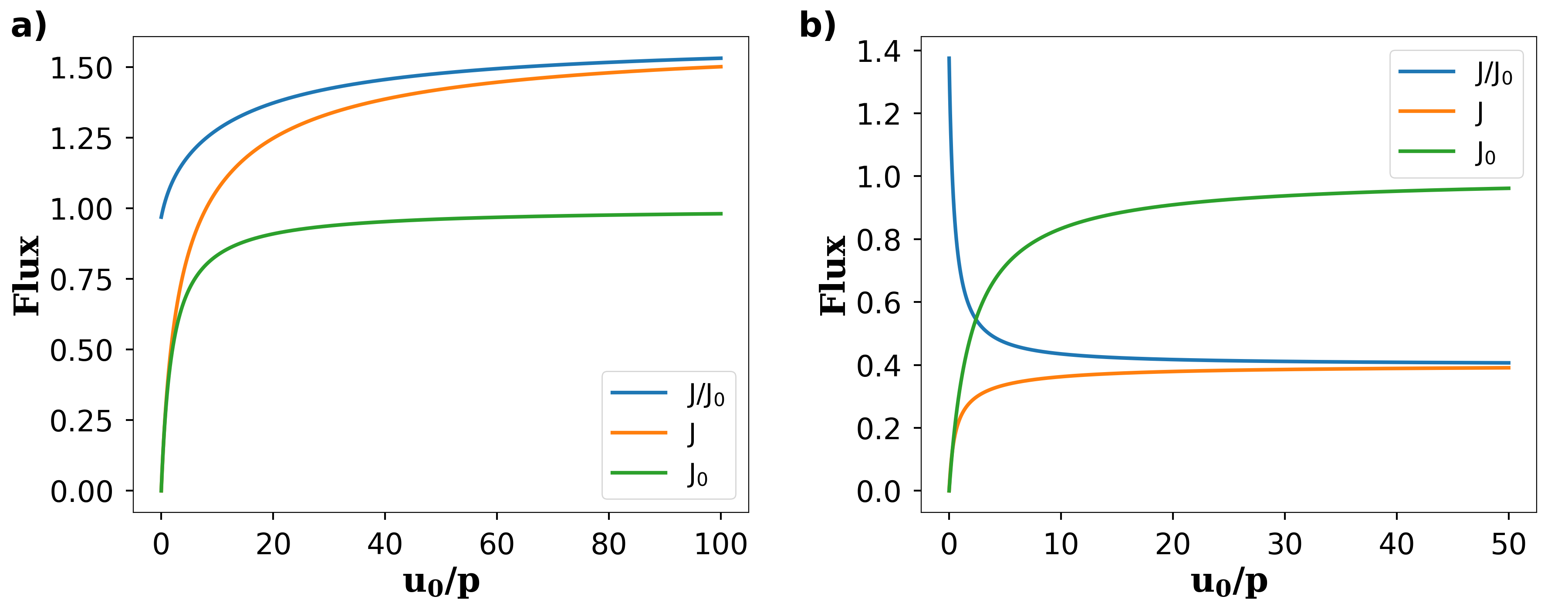}
  \caption{\label{fig:fluxvsc} Plot of currents $J/J_0$, $J$, $J_0$ vs. $u_0/p$ for the for the model of the stochastic gating with changing free-energy. For calculations we used: a) $x=0.5$ and $u_1/p=w_1/p=1$, b) $x=2$ and $u_1/p=w_1/p=1.0$.}
\end{figure*}

Another important factor in the channel-facilitated molecular transport is the concentration gradient $c$ between the entrance and exit from the pore. This is the main driving force to move molecules across the pore. In Appendix~A, we calculated explicitly the derivatives of $J$ (full current) and $J_{n}$ (normalized current) with respect to $c$. Complex behavior is observed because the molecular flux  for the case of no stochastic gating, $J_{0}$, also depends on the concentration gradient. This leads to different behaviors for $J$ and $J_{n}$. Our calculations show that $dJ/dc$ is always positive. This is an expected result because for larger concentration gradients the translocation driving forces are also stronger. At the same time, $dJ_n/dc$ changes sign at $x=1$. It is found (see Appendix~A) that $J_n$ monotonically increases with $c$ for $x<1$  and it decreases for $x>1$. The Figure~\ref{fig:fluxvsc} shows the dependence of currents $J_n$, $J$, $J_0$ on the normalized rate parameter $u_0/p$, which is proportional to $c$, for different $x$ values. One can see that for $x<1$ the normalized current $J_n$ can start below $1$ (see Figure 3a) and increase above one as the concentration increases. While, for $x>1$  (Figure 3b) the opposite can happen. These observations suggest that increasing the concentration gradient can improve the molecular flux in the system with stochastic gating if the new conformation is energetically less favorable ($x<1$). 

\begin{figure*}
  \centering
    \includegraphics[width=0.7\textwidth]{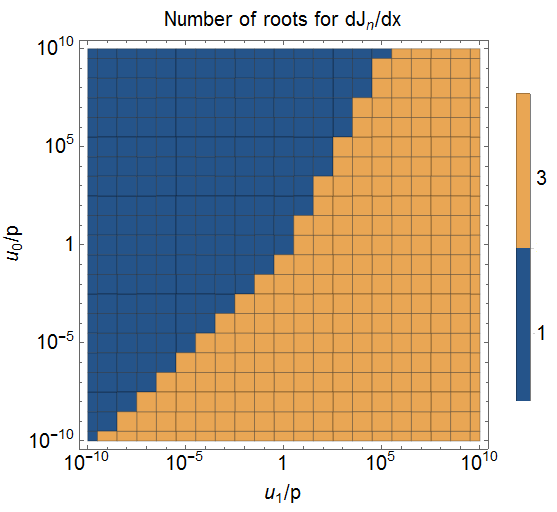}
  \caption{\label{fig:nroots} The map for the number of roots for $dJ_n/dx$ for the model of stochastic gating with changing free-energy profile.}
\end{figure*}

\begin{figure*}
  \centering
   \includegraphics[width=1.0\textwidth]{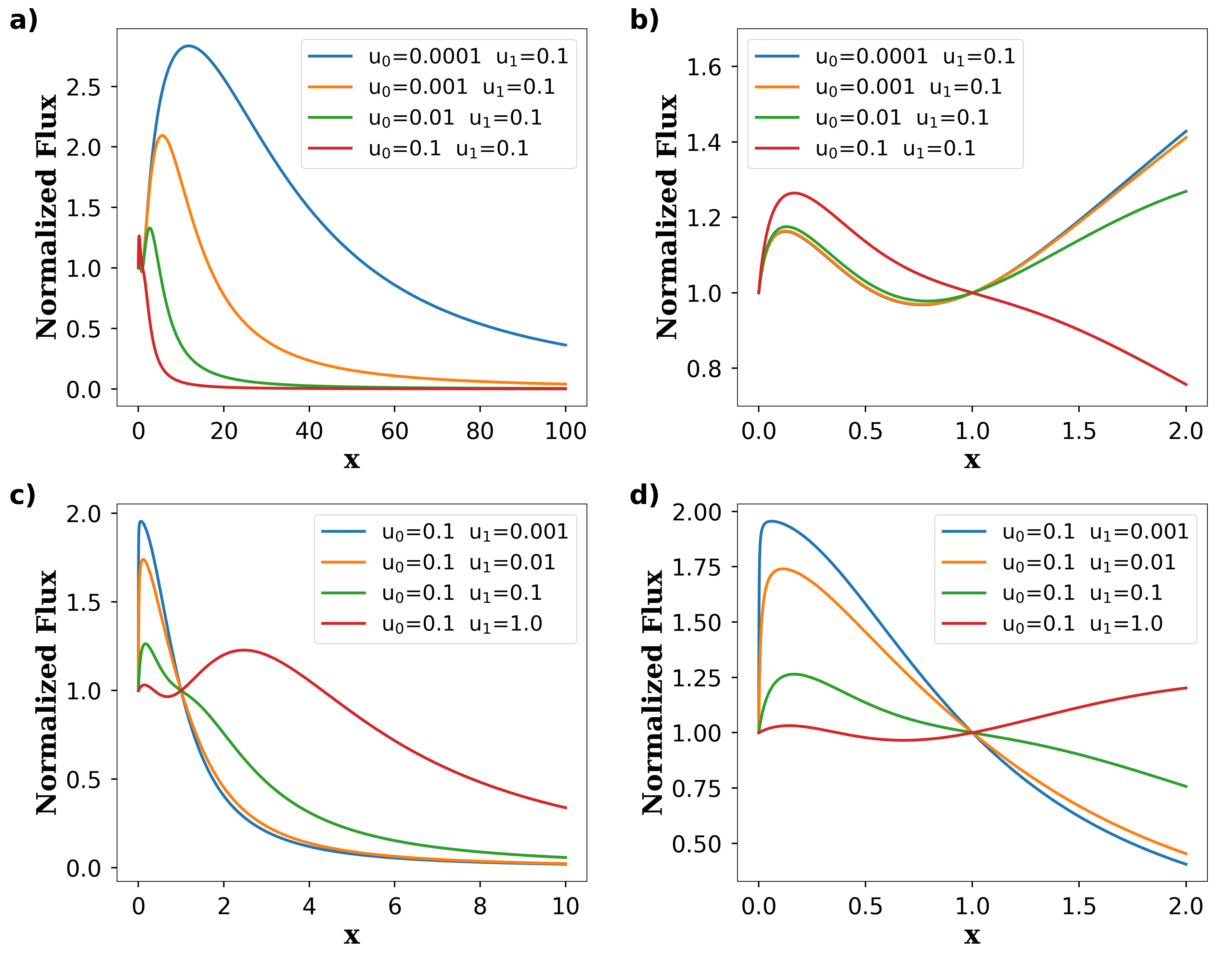}
  \caption{\label{fig:fluxvsx} Normalized molecular fluxes through pores as a function of the interaction energy parameter $x$. For calculations $p=1$ was assumed. }
\end{figure*}

The molecular transport via stochastically fluctuating pores can also be influenced by changing the difference in interaction energy $E$ between two conformations, i.e., by varying the parameter $x=e^{\beta E/2}$. To simplify the analysis, we assume that $w_1=u_1$ and consider the normalized transition rates $u_0/p$ and $u_1/p$ (or equivalently $p=1$). The derivative of the normalized current $J_{n}$ with respect to the variable $x$ is analyzed numerically, and it is found that $dJ_{n}/dx=0$ leads to only one or only three real positive roots, as shown in Figure~\ref{fig:nroots}. It is found that for large entrance transition rates $u_{0}$ the system tends to have a single maximum in the normalized current as a function of $x$ - see also Figure~\ref{fig:fluxvsx}. In this case, entering into the channel is fast, and the rate-limiting step of the whole process is passing and exiting from the pore. It is clear that varying the difference in interactions energies, one could optimize the flux through both conformations. 

A more complex behavior is observed for large transition rates $u_{1}$, which describe exiting from the channel. There are two maxima and one intermediate minimum in the dependence of $J_{n}$ on the parameter $x$ (see Figure~\ref{fig:fluxvsx}). At these conditions, the translocation dynamics is defined by the entrance into the channel and the conformational fluctuations. The interplay between these processes lead to such complex transport dynamics. However, the important conclusion from our calculations is that biological systems might utilize the stochastic gating to improve the molecular transport by varying the interaction energy difference between different conformations.

\subsection{Stochastic Gating with Symmetry Change}

\begin{figure*}
  \centering
    \includegraphics[width=1.0 \textwidth]{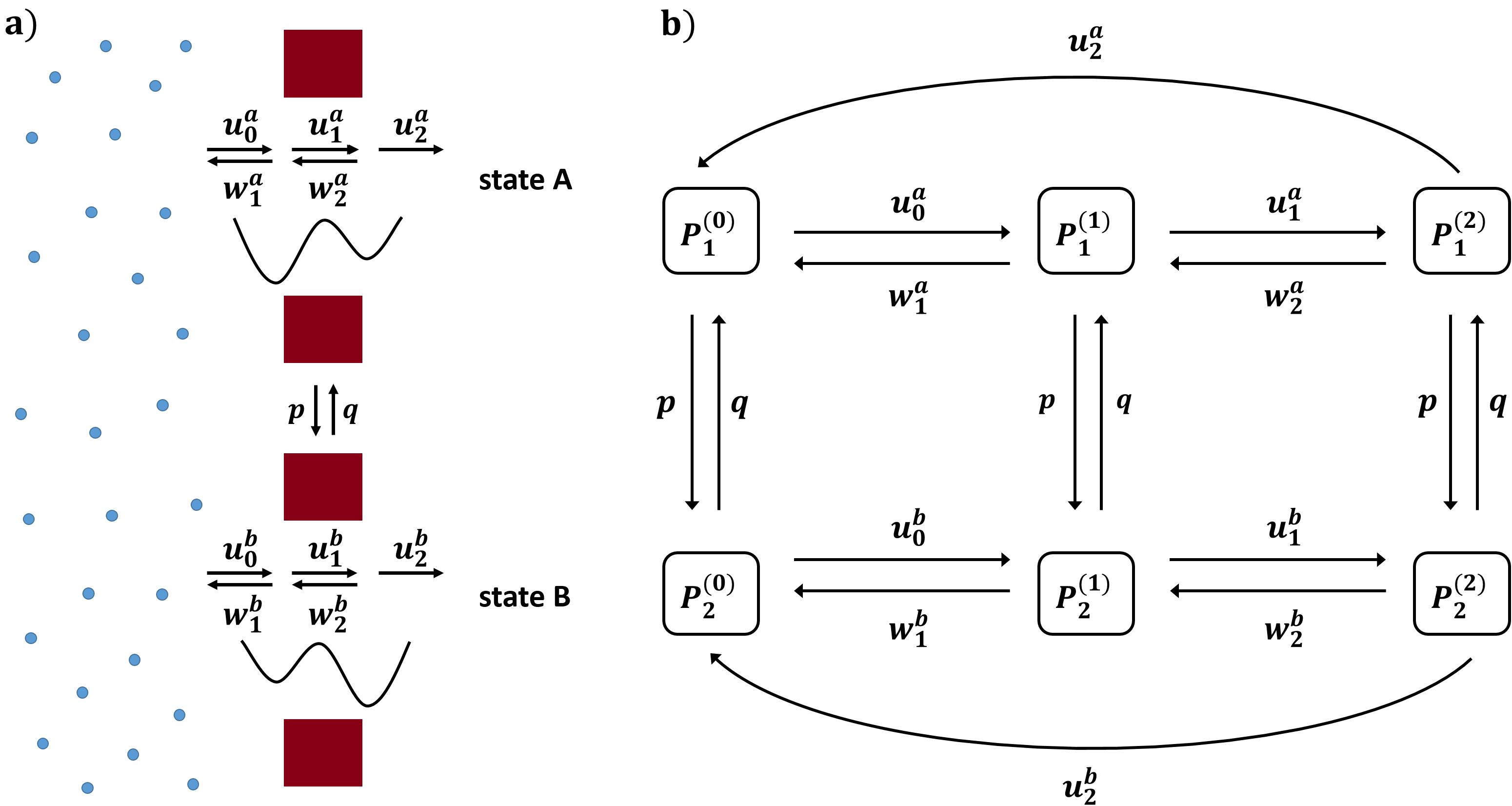}
    \caption{\label{fig:prob2} a) A schematic view of the molecular transport via fluctuating channel in the model of stochastic gating with symmetry changes. b) A corresponding chemical-kinetic diagram for the model.}
\end{figure*}

So far we considered the simplest model of the stochastic gating when there is a single site of interaction between the particle and the pore, and the overall changes in the interactions do not affect the overall symmetry of the free-energy single-well translocation profile. More complex scenarios of the stochastic gating are possible. One of them, which includes a symmetry change for the free-energy double-well translocation profile without varying the average energy of interactions with the pore, is analyzed here. We assume that the channel has two binding sites at which the particle can associate to the pore with different energies as shown in Figure~\ref{fig:prob2}. The interaction potential fluctuates between two states, called $A$ and $B$. In state $A$, the deeper well (where the particle-channel interaction is stronger) is closer to the entrance, while in state $B$ the stronger interacting site is located near the exit. Note that during the translocation the average interaction with the channel is constant, but the shape of the translocation free-energy profile fluctuates between two different double-well potentials.

The possible transitions in the system are shown in Figure~\ref{fig:prob2}. We define $u_{0}^{a}$ and $u_{0}^{b}$ as concentration-dependent entrance rates into the channel in states $A$ and $B$, respectively. The rates $w_{1}^{a}$ and $w_{1}^{b}$ describe the rates of exiting back to the left chamber from the first binding site in the state $A$ and $B$, respectively. The rates $u_{1}^{a}$ and $u_{1}^{b}$ correspond to forward transitions between the first and second binding sites in the state $A$ and $B$, respectively. Similarly, the rates $w_{2}^{a}$ and $w_{2}^{b}$ correspond to backward transitions between the second and first binding sites in the state $A$ and $B$, respectively. Finally, the rates $u_{2}^{a}$ and $u_{2}^{b}$ describe exiting transitions to the right chamber in the state $A$ and $B$, respectively. The system fluctuates between the states A and B with rates $p$ and $q$, respectively (Figure ~\ref{fig:prob2}). We also assume that at the deeper well the particle-pore interaction energy is larger by $\epsilon$ than the interaction energy in the shallow well.

Let us define a probability $P_i^{(j)}(t)$ for the particle to be found in the channel state $i$ ($i=1$ for the state $A$ and $i=2$ for the state $B$) in the particle binding state $j$ ($j=0$ for no bound particle in the pore, $j=1$ for the particle bound in the first site, and $j=2$ for the particle bound in the second well) at time $t$. The temporal evolution of probabilities $P_{i}^{(j)}(t)$ is governed by the following set of master equations:
\begin{equation}
\frac{dP_1^{(0)}(t)}{dt}=-(u_0^a + p) P_1^{(0)}(t) + w_1^a P_1^{(1)}(t) + u_2^a  P_1^{(2)}(t) + q P_2^{(0)}(t)
\label{eq:p2mastp10},
\end{equation}
\begin{equation}
\frac{dP_1^{(1)}(t)}{dt}=-(u_1^a + w_1^a + p) P_1^{(1)}(t) + u_0^a P_1^{(0)}(t) + w_2^a P_1^{(2)} + q P_2^{(1)}(t)
\label{eq:p2mastp11},
\end{equation}
\begin{equation}
\frac{dP_1^{(2)}(t)}{dt}=-(u_2^a + w_2^a + p) P_1^{(2)}(t) + u_1^a P_1^{(1)}(t) + q P_2^{(2)}(t)
\label{eq:p2mastp12},
\end{equation}
\begin{equation}
\frac{dP_2^{(0)}(t)}{dt}=-(u_0^b + q) P_2^{(0)}(t) + w_1^b P_2^{(1)}(t) + u_2^b  P_2^{(2)}(t) + p P_1^{(0)}(t)
\label{eq:p2mastp20},
\end{equation}
\begin{equation}
\frac{dP_2^{(1)}(t)}{dt}=-(u_1^b + w_1^b + q) P_2^{(1)}(t) + u_0^b P_2^{(0)}(t) + w_2^b P_2^{(2)} + p P_1^{(1)}(t)
\label{eq:p2mastp21},
\end{equation}
\begin{equation}
\frac{dP_2^{(2)}(t)}{dt}=-(u_2^b + w_2^b + q) P_2^{(2)}(t) + u_1^b P_2^{(1)}(t) + p P_1^{(2)}(t)
\label{eq:p2mastp22}.
\end{equation}
In addition, $P_i^{(j)}(t)$ must satisfy the following normalization condition:
\begin{equation}
P_1^{(0)}(t) + P_1^{(1)}(t) + P_1^{(2)}(t) + P_2^{(0)}(t) + P_2^{(1)}(t) + P_2^{(2)}(t) = 1.
\label{eq:p2pnorm}
\end{equation}

Again, we are interested in the stationary-state solutions when $dP_i^{(j)}/dt=0$. In that case, the system of Equations (\ref{eq:p2mastp10}-\ref{eq:p2pnorm}) can be solved analytically and the general solution is given in Equations (\ref{eq:gensolp10}-\ref{eq:gensolp22}) in Appendix~B. Now we can explicitly estimate the molecular flux through the channel via
\begin{equation}
J = u_2^a P_1^{(2)} + u_2^b P_2^{(2)}
\label{eq:fluxdefp2}.
\end{equation}

The general expression for the particle current $J$ is presented in Equation (\ref{eq:fluxgen}) in Appendix B. One can see that the equation for molecular flux is symmetric with respect to $A \leftrightarrow B$ and $p \leftrightarrow q$ transformations, as expected.  But to understand better the dynamic behavior of the system, we simplify this expression by making several simple assumptions. First of all, we can take into account the detailed-balance-like arguments for the transitions rates. One can write:
\begin{equation}
\frac{u_0^b}{w_1^b} = \frac{u_0^a}{w_1^a} e^{-\beta \varepsilon}, \ \ \  \text{and} \ \ \  \frac{u_1^b}{w_2^b} = \frac{u_1^a}{w_2^a} e^{\beta \varepsilon}
\label{eq:detbal}.
\end{equation}
where $\varepsilon \ge 0$ is the difference in the interaction energies when particle is found in different binding sites in the pore. The physical meaning of these expressions is easy to interpret: the particle enters faster to the sites with lower energy and it exits slower from these sites, while entrance to the higher-energy sites is slower and the exit from them is faster.

Using Equation~(\ref{eq:detbal}), we can explicitly rewrite transition rates as
\begin{equation}
u_0^b = u_0^a e^{\beta (\theta - 1) \varepsilon}, \ u_1^b = u_1^a e^{\beta \theta \varepsilon}, \  w_1^b = w_1^a e^{\beta \theta \varepsilon}, \ w_2^b = w_2^a e^{\beta (\theta - 1) \varepsilon}; 
\end{equation}
where the parameter $\theta$, $0 \leqslant \theta \leqslant 1$, describes how the interaction energy difference $\varepsilon$ influences the forward and backward transition rates. In addition, it is assumed that $u_2^b = u_2^a e^{-\beta \varepsilon}$, which again reflects the fact that it is more difficult to exit from the site with stronger interactions. For simplicity, we also take that $p=q$ and $\theta=1/2$. This is equivalent to the following assumption: $u_1^b=w_2^a$ and $u_1^a=w_2^b$ (see Figure~\ref{fig:prob2}). If we define $u_0^a \equiv u_0$, $u_1^a \equiv u_1$, $u_2^a \equiv u_2$, $w_1^a \equiv w_1$, $w_2^a \equiv w_2$ and $x \equiv e^{\beta \varepsilon/2}$, then the transition rates can be presented as $u_0^b = u_0 x^{-1}$, $u_1^b = u_1 x$, $u_2^b = u_2 x^{-2}$, $w_1^b = w_1 x$, $w_2^b = w_2 x^{-1}$. All these simplifications lead to the following expression for the molecular flux, 
\begin{widetext}
\begin{equation}
J = \frac{u_0 u_1 u_2}{2} \frac{p^2 (1+x)^2 (1+x^2) + p(1+x)(x \alpha + \beta(x) + u_0 (x-1)^2) + x \gamma + \delta(x)}{p^2 (1+x)(x^2 \gamma + \delta(x) + x(x-1)^2 u_2(u_1 + w_1)) + p(\alpha \delta(x) + x \beta(x)\gamma) + \gamma \delta(x)}
\label{eq:fluxsimp1p2},
\end{equation}
\end{widetext}
where
\begin{small}
\begin{eqnarray}
\alpha &=& u_0 + u_1 + u_2 + w_1 + w_2, \\
\beta(x) &=& u_0 x + u_1 x^3 + u_2 + w_1 x^3 + w_2 x, \\
\gamma &=& u_0 u_1 + u_0 u_2 + u_0 w_2 + u_1 u_2 + u_2  w_1 + w_1 w_2, \\
\delta(x) &=& u_0 u_1 x^3 + u_0 u_2 + u_0 w_2 x + u_1 u_2 x^2 + u_2  w_1 x^2 + w_1 w_2 x^3
\label{eq:fluxsimp1var}.
\end{eqnarray}
\end{small}

Because of the symmetry between states $A$ and $B$, changing the sign of the interaction energy difference, i.e., $\epsilon \rightarrow -\epsilon$, is identical to $x \rightarrow 1/x$, $u_{0} \rightarrow u_{0}/x$, $u_{1} \rightarrow u_{1}x$, $w_{1} \rightarrow w_{1}x$, $w_{2} \rightarrow w_{2}/x$ and $u_{2} \rightarrow u_{2}/x^{2}$. Under these transformations, the expression for the current given in Eq. (\ref{eq:fluxsimp1var}) does not change, and this means that we can consider only positive $\epsilon$ ($x \ge 1$) to analyze the molecular transport through fluctuating pores. For $x=1$ ($\varepsilon=0$), the Equation (\ref{eq:fluxsimp1p2}) gives
\begin{equation}
J = \frac{u_0 u_1 u_2}{u_0 (u_1 + u_2 + w_2) + u_1 u_2 + w_1 (u_2 + w_2)}
\label{eq:fluxx1}.
\end{equation}
In this case, the particle at both binding sites always have the same interactions with the channel and there is no symmetry fluctuations in the system. In the stationary-state limit, the problem is analogous to a single random walker moving on infinite  three-state periodic lattice (corresponding to two binding sites in the pore and the state outside of the pore), which has been widely explored in the literature.\cite{derrida1983velocity,Kolomeisky2007,kolomeisky2007molecular} For $x \to \infty$ ($\varepsilon \to \infty$), the Equation (\ref{eq:fluxsimp1p2}) gives
\begin{equation}
J = \frac{u_0 u_1 u_2 (p + u_1 + w_1)}{2\left[p (u_1(u_0 + u_2) + w_1 (u_2 + w_2)) + (u_1 + w_1) \gamma\right]}
\label{eq:fluxxinf}.
\end{equation}

It is interesting to analyze the translocation dynamics in several limiting cases. For $u_0 \gg 1$ (fast entrance rates), the Equation (\ref{eq:fluxsimp1p2}) simplifies into
\begin{eqnarray}
J = \frac{u_1 u_2}{2} \times &  & \nonumber \\ & \frac{p(1+x)(1+x^2) + u_1 x(1+x^2) + u_2 (1+x) + 2 w_2 x}{p(u_1 + u_2 + w_2) x^2 + (p + u_1 + u_2 + w_2)(u_1 x^3 + u_2 + w_2 x)}  &
\label{eq:fluxu0gg}.
\end{eqnarray}
One can see that in this case the flux is independent of the rate $w_1$. This is because the particle that returned to the left chamber is immediately introduced back into the channel. 

For $u_1 \gg 1$, which corresponds to fast forward transitions inside the channel from the first to the second binding sites, the Equation (\ref{eq:fluxsimp1p2}) reduces to
\begin{equation}
J = \frac{u_0 u_2}{2}\frac{p(1+x)(1+x^2) + u_0 (1+x^2) + u_2 (1+x)}{x(p(u_0 + u_2) x^2 + (p + u_0 + u_2)(u_0 x + u_2))}
\label{eq:fluxu1gg}.
\end{equation}
Note that in this case the molecular flux $J$ does not depend on either $w_1$ or $w_2$ transition rates. This physically means that the rate of the transition from the first binding site to the second one is so fast that the system does not have time to exit back to the left chamber or to move backward from the second binding site. 

For $u_2 \gg 1$, which describes fast rates to exit the channel to the right chamber, from the Equation (\ref{eq:fluxsimp1p2}) we obtain
\begin{small}
\begin{equation}
J = \frac{u_0 u_1}{2}\frac{(1+x)(p(1+x) + u_0 + (u_1 + w_1) x)}{p(1+x)(u_0 + (u_1 + w_1)x) + (u_0 + u_1 + w_1)(u_0 + (u_1 + w_1)x^2)}
\label{eq:fluxu2gg}.
\end{equation}
\end{small}
In this case, we observe that the particle current is independent of the backward transition rate $w_{2}$. This can be also easily understood because as soon as the particle reaches the second binding site, it immediately exits to the right, and the probability of the backward transition inside the channel is negligible.

Now let us discuss the behavior of molecular flux when conformational fluctuations rates, concentration gradients and interaction energies are varied. We will do this by investigating the derivatives of current $J$ with respect to corresponding variables. But to simplify our calculations even further, we will make the following additional assumptions on transitions rates: $w_1=u_1$, and $w_2=u_2=u_1 x^2$. This corresponds to a physically reasonable situation when the transition states for all transformations have the same energy: see Figure~\ref{fig:prob2}. The details of calculations are presented in Appendix C.

The molecular transport depends on the frequency of conformational transitions. The results of our calculations are presented in Figure 7. It can be shown that there is a special interaction energy $\epsilon_e$ ($x_{e}=e^{\beta \epsilon_{e}/2}$) such that for stronger interactions ($x \ge x_{e}$) there is always a minimum in the molecular flux as a function of the conformational frequency change.  This means that increasing the frequency of fluctuations first lowers the molecular flux, but after passing the critical conformational transition rate the molecular flux starts to increase. However, there is also a parameter range when the particle current will always increase with increasing the frequency of conformational transitions. These observations can be explained using the following arguments. For large interaction energy differences between the binding sites ($x>x_{e}$), increasing first the frequency of conformational transitions will lower the molecular flux because the system will spend most of the time by being trapped in the deepest wells of the states $A$ and $B$ instead of trying to pass the channel. But eventually for larger $p$ this effect will be less important since increasing the frequency of conformational fluctuations will decrease the trapping of the molecules at the strongly interacting sites of the pore. For smaller interaction differences ($x<x_{e}$), only the untrapping effect will play the role. Note that the dependence of the molecular flux on frequency of conformational fluctuations in the model of stochastic gating with symmetry changes is different from the model with free-energy variations. This shows the role of symmetry variations in the stochastic gating phenomena.

\begin{figure*}
  \centering
    \includegraphics[width=1.0\textwidth]{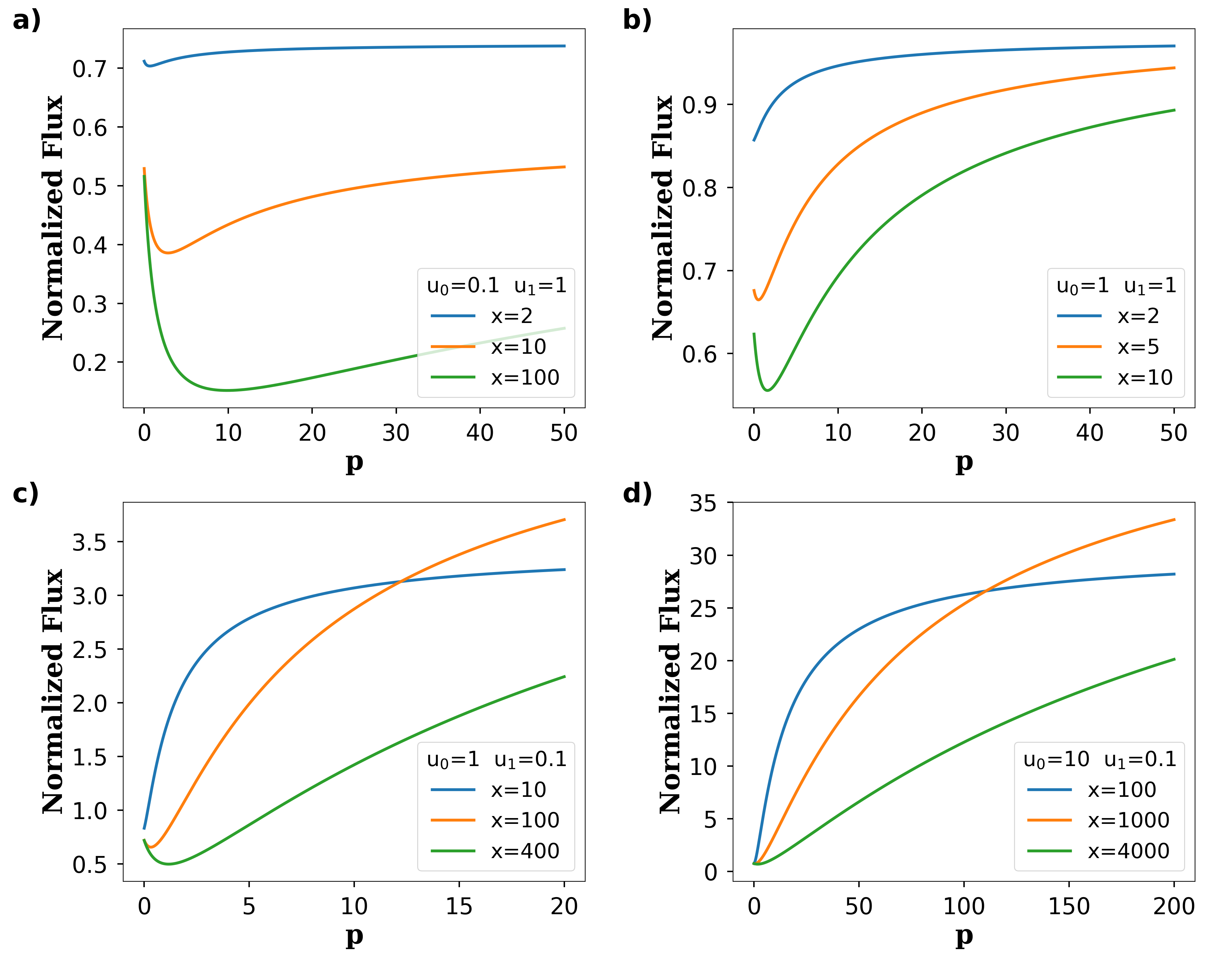}
  \caption{\label{fig:p2fluxvsp} Normalized molecular fluxes as a function of the conformational transition rate $p$ for the model of the stochastic gating with changing symmetry. The following parameters are utilized in calculations: a) $u_0=0.1$ and $u_1=1$ (for these values $x_e=1$); b) $u_0=1$ and $u_1=1$ ($x_e=3.10$); c) $u_0=1$ and $u_1=0.1$ ($x_e=24.83$); and d) $u_0=10$ and $u_1=0.1$ ($x_e=242.10$).}
\end{figure*}


The entrance rate $u_0$ is proportional to the  concentration of the particles on the left side of the channel: $u_0=c k_{on}$. Thus, considering the dependence of the molecular flux on the parameter $u_{0}$ gives the effect of the concentration gradient on the particle current. Our explicit calculations (Appendix C)  show that  $dJ/d u_0 >0$ for any $u_0>0$, $u_1>0$, as expected, since the concentration gradient is the main driving force for the molecular transport across the channel. This means that increasing the concentration gradient will always improve the molecular transport via pores.


The dependence of the molecular fluxes on the interaction energy difference is more complex, as shown in Figure 8. There are situations when increasing the interaction energy difference always lowers the molecular flux via the pores. At another range of parameters, changing $x$ might actually lead to the non-monotonic behavior, with only one maximum or with minimum and maximum - see Figure 8. These observations are the result of several competing processes. While the system is in the conformation $B$, increasing $\epsilon$ will stimulate the molecule to translocate to the second site, which increases the flux. However, when the interaction difference becomes very large, the molecule can be trapped at the sites with strongest interactions. If the system is in the state $A$, increasing the interaction energy difference will only trap the molecule in the deepest well without moving it forward.

Our theoretical analysis presents a very rich dynamic behavior for the systems with stochastic gating. The molecular translocation via channels can be influenced by modifying the frequency of conformational changes, the molecule/pore  interaction energies and the concentration gradients. It seems reasonable to suggest that the nature has multiple tools to tune the channel transport to fulfill the necessary biological functions.

\begin{figure*}
  \centering
    \includegraphics[width=1.0\textwidth]{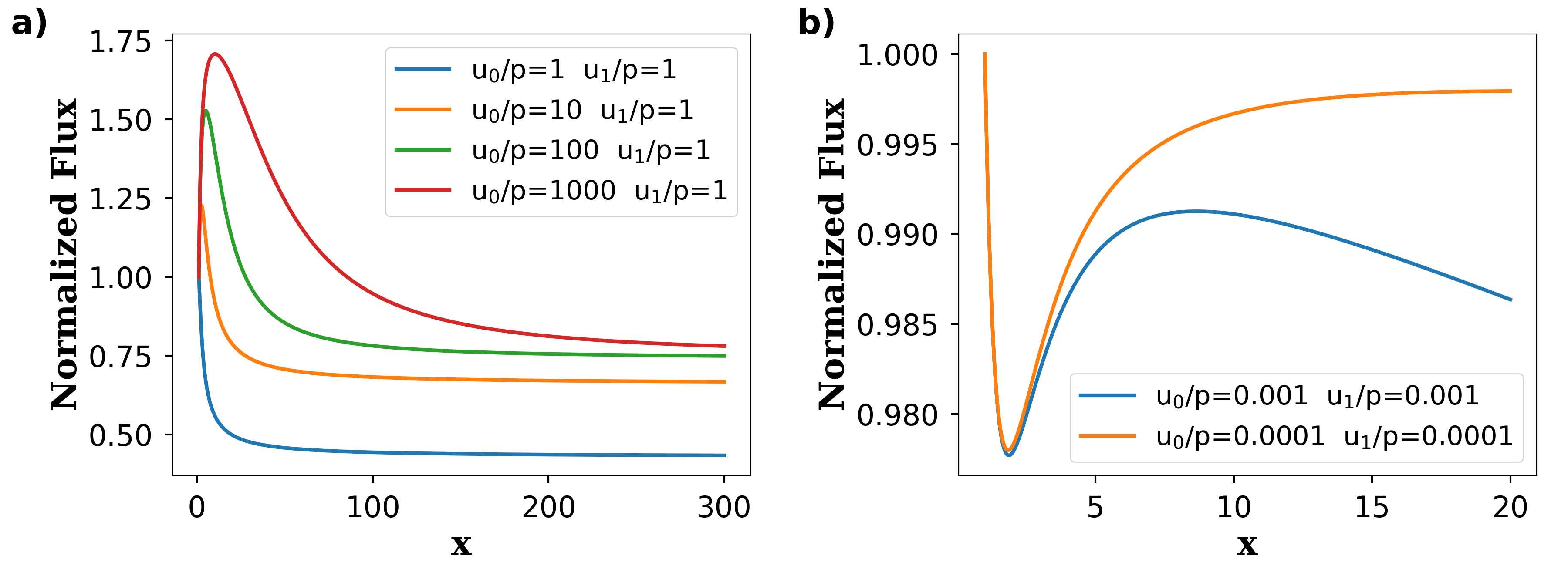}
  \caption{\label{fig:p2fluxvsx3} Plot of $J/J_0$ vs. $x$ for for the model of the stochastic gating with changing symmetry of the channel.}
\end{figure*} 


\section{\label{sec:conc} Summary and Conclusions}

We developed a discrete-state chemical-kinetic approach to investigate the effect of stochastic gating in the channel-facilitated molecular transport. Our theoretical analysis explicitly evaluates the particle currents through the pores in terms of transition rates between various chemical states and conformation. It allows us to specifically investigate two different models of stochastic gating. In the first model, the stochastic gating leads to the changes in the translocation free-energy profile but without symmetry variations. It is found that increasing the frequency of conformational transitions and the concentration gradients between different parts of the channel will always increase the particle current through the system. At the same time, varying the interaction energy between the molecules and the pores generally leads to non-monotonic behavior. A more complex dynamic behavior is observed in the second model of stochastic gating that involves symmetry variations in the free-energy translocation profile without changing the overall interactions. While increasing the concentration gradient will always accelerate the molecular fluxes, the dependence on the frequency of conformational fluctuations and on interaction energies is non-monotonic. We presented microscopic arguments to explain these observations. Importantly, in both models we do not observe phenomena similar to the stochastic resonance when there is an optimal rate of conformational transitions that leads to a maximal particle current.

Although our theoretical method is able to quantitatively describe stochastic gating phenomena, it is important to note that our approach is rather very  simplified and many realistic features are not taken into account. It is clear that real biological system will be very different from simplified models considered in this work. More complex free-energy translocation profiles and multiple conformational transitions are expected in biological cells. In addition, in our approach it was assumed that stochastic gating is taking place at the stationary conditions, but it is not guaranteed that biological systems can satisfy this. Despite these limitations, our theoretical approach provides a fully quantitative molecular picture of complex processes associated with stochastic gating that might be utilized for the development of more advances theoretical descriptions. It might be also useful in analyzing experimental observations related to biological transport processes.

\section*{Acknowledgments}

The work was supported by the Welch Foundation (C-1559), by the NSF (CHE-1664218), and by the Center for Theoretical Biological Physics sponsored by the NSF (PHY-1427654).

\begin{widetext}

\appendix
\setcounter{equation}{0}
\renewcommand{\theequation}{\thesection\arabic{equation}}

\section{Derivation of the particle flux for the model  with changing free-energy profile}

Here we explicitly calculate the derivatives of the particle flux with respect to different parameters. The change of the current with respect to the parameter $p$ is given by 
\begin{eqnarray}
\frac{dJ}{dp} &=& \frac{u_0 u_1 x}{1+x^2}\frac{(W-U)(U x^3 - W)}{(p x (U+W) + U W)^2} \nonumber
\\
&=& \frac{u_0 u_1}{1 + x^2}\frac{x (1 - x)^2 (u_0 (1 + x + x^2) + w_1)(u_1 (1 + x + x^2) + w_1 x(x+1))}{(p x (U+W) + U W)^2}\geqslant 0
\label{eq:fluxdp},
\end{eqnarray}
while the dependence on the concentration gradient is equal to
\begin{eqnarray}
\frac{dJ}{dc} &=& \frac{k_{on} u_1}{1+x^2}\frac{2 p^2 x^2 (1+x^3) (U_1 + W_1) + p x [(U x^3 + W)(U_1 + W_1) + 2 (x^3 U W_1 + W U_1)] +}{(p x (U+W) + U W)^2} \nonumber
\\
&& \frac{ + x^3 U^2 W_1 + W^2 U_1}{} \nonumber\geqslant 0
\label{eq:fluxdc},
\end{eqnarray}
where
\begin{equation}
U_1 =  u_1 + w_1, \ \ \ W_1 =  u_1 + w_1 x
\label{eq:UWsimp2}.
\end{equation}
The expression is different for the normalized particle current,
\begin{eqnarray}
\frac{d}{dc}\left( \frac{J}{J_0} \right) &=& \frac{k_{on} x}{1+x^2}\frac{(W - U x^3)(2 p^2 x(1+x^3) + p(W + 3 x^3 U) + U^2 x^2)}{(p x (U+W) + U W)^2} \nonumber \\
&=& \frac{k_{on} x (1 - x)}{1+x^2}\frac{(u_1(1+x+x^2) + w_1 x(1+x))(2 p^2 x(1+x^3) + p(W + 3 x^3 U) + U^2 x^2)}{(p x (U+W) + U W)^2} 
\label{eq:normfluxdc}.
\end{eqnarray}
We can also write the normalized current in the following way,
\begin{eqnarray}
\frac{J}{J_0}(p u_0, p u_1, p w_1) = \frac{U}{1+x^2} \frac{2 x (1+x^3) + U x^3 + W}{x(U + W) + U W}
\label{eq:normflux}.
\end{eqnarray}


\section{Stationary solutions for the model with changing symmetry}

Here we present exact expressions for the solution of the system of Equations (\ref{eq:p2mastp10}-\ref{eq:p2pnorm}):
\begin{equation}
P_1^{(0)} = \frac{q}{p+q} \frac{(q^2 + q \alpha_b + \gamma_b) \zeta_a + p^2 \zeta_b + p(u_0^b \zeta_{ba} + (\alpha_a - u_0^a) \zeta_b)+ p q (\zeta_{ab} + \zeta_{ba})}{q(q + \alpha_b)\gamma_a + p(p + \alpha_a)\gamma_b + pq (\gamma_{ab} + \gamma_{ba}) + \gamma_a \gamma_b}
\label{eq:gensolp10},
\end{equation}
\begin{equation}
P_1^{(1)} = \frac{q}{p+q} \frac{((q^2 + q \alpha_b + \gamma_b)u_0^a + p q u_0^b)(u_2^a + w_2^a) + (p^2 u_0^b + p q u_0^a)(u_2^b + w_2^b) + p \eta_{ab}}{q(q + \alpha_b)\gamma_a + p(p + \alpha_a)\gamma_b + pq (\gamma_{ab} + \gamma_{ba}) + \gamma_a \gamma_b}
\label{eq:gensolp11},
\end{equation}
\begin{equation}
P_1^{(2)} = \frac{q}{p+q} \frac{(q^2 + q \alpha_b + \gamma_b) u_0^a u_1^a + p^2 u_0^b u_1^b + p q (u_0^b u_1^a + u_0^a u_1^b) + p u_0^b \lambda_{ab}}{q(q + \alpha_b)\gamma_a + p(p + \alpha_a)\gamma_b + pq (\gamma_{ab} + \gamma_{ba}) + \gamma_a \gamma_b}
\label{eq:gensolp12},
\end{equation}
\begin{equation}
P_2^{(0)} = \frac{p}{p+q} \frac{q^2 \zeta_a + (p^2 + p \alpha_a + \gamma_a)\zeta_b + q(u_0^a \zeta_{ab} + (\alpha_b - u_0^b) \zeta_a) + p q (\zeta_{ab} + \zeta_{ba})}{q(q + \alpha_b)\gamma_a + p(p + \alpha_a)\gamma_b + pq (\gamma_{ab} + \gamma_{ba}) + \gamma_a \gamma_b}
\label{eq:gensolp20},
\end{equation}
\begin{equation}
P_2^{(1)} = \frac{p}{p+q} \frac{((p^2 + p \alpha_a + \gamma_a)u_0^b + p q u_0^a)(u_2^b + w_2^b) + (q^2 u_0^a + p q u_0^b)(u_2^a + w_2^a) + q \eta_{ba}}{q(q + \alpha_b)\gamma_a + p(p + \alpha_a)\gamma_b + pq (\gamma_{ab} + \gamma_{ba}) + \gamma_a \gamma_b}
\label{eq:gensolp21},
\end{equation}
\begin{equation}
P_2^{(2)} = \frac{p}{p+q} \frac{q^2 u_0^a u_1^a + (p^2 + p \alpha_a + \gamma_a) u_0^b u_1^b + p q (u_0^b u_1^a + u_0^a u_1^b) + q u_0^a \lambda_{ba}}{q(q + \alpha_b)\gamma_a + p(p + \alpha_a)\gamma_b + pq (\gamma_{ab} + \gamma_{ba}) + \gamma_a \gamma_b}
\label{eq:gensolp22},
\end{equation}
where
\begin{eqnarray}
\alpha_a &=& u_0^a + u_1^a + u_2^a + w_1^a + w_2^a \\
\alpha_b &=& u_0^b + u_1^b + u_2^b + w_1^b + w_2^b \\
\gamma_a &=& u_0^a (u_1^a + u_2^a + w_2^a) + u_1^a u_2^a + w_1^a (u_2^a + w_2^a) \\
\gamma_b &=& u_0^b (u_1^b + u_2^b + w_2^b) + u_1^b u_2^b + w_1^b (u_2^b + w_2^b) \\
\gamma_{ab} &=& u_0^a (u_1^b + u_2^b + w_2^b) + u_1^a u_2^b + w_1^a (u_2^b + w_2^b) \\
\gamma_{ba} &=& u_0^b (u_1^a + u_2^a + w_2^a) + u_1^b u_2^a + w_1^b (u_2^a + w_2^a) \\
\lambda_{ab} &=& u_0^a u_1^b + u_1^a (u_1^b + + u_2^b + w_2^b) + w_1^a u_1^b \\
\lambda_{ba} &=& u_0^b u_1^a + u_1^b (u_1^a + + u_2^a + w_2^a) + w_1^b u_1^a \\
\zeta_a &=& u_1^a u_2^a + w_1^a (u_2^a + w_2^a) \\
\zeta_b &=& u_1^b u_2^b + w_1^b (u_2^b + w_2^b) \\
\zeta_{ab} &=& u_1^a u_2^b + w_1^b (u_2^a + w_2^a) \\
\zeta_{ba} &=& u_1^b u_2^a + w_1^a (u_2^b + w_2^b) \\
\eta_{ab} &=& (u_0^a(u_0^b + w_1^b) + u_0^b (u_2^a + w_2^a)) (u_2^b + w_2^b) + u_1^b (u_0^a u_2^b + u_0^b w_2^a) \\
\eta_{ba} &=& (u_0^b(u_0^a + w_1^a) + u_0^a (u_2^b + w_2^b)) (u_2^a + w_2^a) + u_1^a (u_0^b u_2^a + u_0^a w_2^b)
\label{eq:gensolsup}.
\end{eqnarray}

Given this solution, we calculate the particle current through the channel from the Equation (\ref{eq:fluxdefp2}),
\begin{small}
\begin{equation}
J = \frac{q(q^2 + q (p + \alpha_b) + \gamma_b) u_0^a u_1^a + p(p^2 + p (q + \alpha_a) + \gamma_a) u_0^b u_1^b + p q(p + q)(u_0^b u_1^a + u_0^a u_1^b) + p q (u_0^a \lambda_{ba} + u_0^b \lambda_{ab})}{(p+q)(q(q + \alpha_b)\gamma_a + p(p + \alpha_a)\gamma_b + pq (\gamma_{ab} + \gamma_{ba}) + \gamma_a \gamma_b)}
\label{eq:fluxgen}.
\end{equation}
\end{small}


\section{Molecular flux for the model of stochastic gating with symmetry fluctuations}

From the Equation (\ref{eq:fluxsimp1p2}), the following expression can be obtained for the particle flux $J$ when $w_1=u_1$, $w_2=u_2$, and $u_2=x^2 u_1$.
\begin{equation}
J = \frac{A}{B}
\label{eq:fluxsimp2p2},
\end{equation}
where
\begin{small}
\begin{eqnarray}
A &=& u_0 u_1 [p^2 (1+x)^2 (1+x^2) + p(1+x)(u_0 (1+x^2) + u_1 x (2 + x + 5 x^2)) + u_1 x (u_1 x^2 (3 + 2 x + x^2) + u_0 (1 + x + 4 x^2)]
\label{eq:fluxsimp2p2A},
\end{eqnarray}
\end{small}
and
\begin{small}
\begin{eqnarray}
B &=& 2[p^2 (1+x)(2 u_0 (1 + x + x^2) + u_1 x (2 + x + 3 x^2)) + \nonumber \\
&+& p(2 u_0^2 (1 + x + x^2) + u_0 u_1 (2 + x + 2 x^2)(1 + 2 x + 3 x^2) +  u_1^2 x^2 (4 + 5 x + 13 x^2 + 2 x^3)) + \\
&+& u_1 (u_0 (1 + 2 x^2) + 3 u_1 x^2)(u_0 (1 + 2 x) + u_1 x^2 (2 + x))] \nonumber
\label{eq:fluxsimp2p2B}.
\end{eqnarray}
\end{small}


Given Equation (\ref{eq:fluxsimp2p2}), we can calculate the first derivative of the current with respect to the variable $p$,
\begin{equation}
\frac{dJ}{dp} = \frac{A_{1}}{B_{1}}
\label{eq:fluxdpsimp},
\end{equation}
where
\begin{small}
\begin{eqnarray}
A_{1} &=& u_0 u_1^2(x-1)^2 [p^2 (1+x)^2 (u_0 (2 + 3 x + 9 x^2 + 6 x^3 + 6 x^4) + u_1 x^3 (1 + 2x + 2x^2)) + \nonumber \\
&+& 2p(1+x) (u_0^2 (1 + x +2 x^2)(1 + 2 x + 2 x^2) + u_0 u_1 x^2 (3 + 9 x + 9 x^2 + 8 x^3 + 2 x^4) + 2 u_1^2 x^5) + \\
&+& u_0^3 (1 + x + 2 x^2)(1 + 2 x + 2 x^2) + u_0^2 u_1 x^2 (5 + 9 x + 13 x^2 + 4 x^3 + 2 x^4) + u_0 u_1^2 x^4 (7 + 6 x +16 x^2 + 4 x^3) +   \nonumber \\
&+& u_1^3 x^6 (1 - 6 x - 2 x^2)] \nonumber 
\label{eq:fluxdpsimpA},
\end{eqnarray}
\end{small}
and
\begin{small}
\begin{eqnarray}
B_{1} &=& 2[p^2 (1+x)(2 u_0 (1 + x + x^2) + u_1 x (2 + x + 3 x^2)) + \nonumber \\
&+& p(2 u_0^2 (1 + x + x^2) + u_0 u_1 (2 + x + 2 x^2)(1 + 2 x + 3 x^2) +  u_1^2 x^2 (4 + 5 x + 13 x^2 + 2 x^3)) + \\
&+& u_1 (u_0 (1 + 2 x^2) + 3 u_1 x^2)(u_0 (1 + 2 x) + u_1 x^2 (2 + x))]^2  \nonumber
\label{eq:fluxdpsimpB}.
\end{eqnarray}
\end{small}


The expression above shows that $dJ/dp$ can have up to two roots, where one of them will always be negative, i.e., unphysical for our model. Indeed, the numerator of $dJ/dp$ has a form $a p^2 + b p + c$, where $a$ and $b$ are always positive for $u_0>0$, $u_1>0$ and $x>1$. Thus, the positive solution will only exist when $c<0$, which can be satisfied either for $u_1/u_0$ or $x$ being sufficiently large. In other words, we conclude that at large enough $u_1/u_0$ the solution for $dJ/dp=0$ exists for the whole range of $x \in (1, \infty)$, and otherwise it exists above the certain value of $x$ ($x>x_e$, where $x_e>1$).

The second derivative of the current with respect to $p$ at the extremum will have the following form,
\begin{equation}
\left(\frac{d^2 J}{dp^2}\right)_{p=p_0} = \frac{A_{2}}{B_{2}}
\label{eq:fluxdp2simp},
\end{equation}
where
\begin{small}
\begin{eqnarray}
A_{2} &=& 2 u_0 u_1^4 x^4 (x-1)^2 (x+1) (p_0(1+x) (u0 (2 + 3 x + 9 x^2 + 6 x^3 + 6 x^4) + u1 x^3 (1 + 2 x + 2 x^2)) + \nonumber \\
&+& u_0^2 (1 + x + 2 x^2)(1 + 2 x + 2 x^2) + 2 u_1^2 x^5 + u_0 u_1 x^2 (3  + 9 x + 9 x^2 + 8 x^3 + 2 x^4))
\label{eq:fluxdp2simpA},
\end{eqnarray}
\end{small}
and
\begin{small}
\begin{eqnarray}
B_{2} &=& (p_0^2 (1+x)(2 u_0 (1 + x + x^2) + u_1 x (2 + x + 3 x^2)) + \nonumber \\
&+& p_0(2 u_0^2 (1 + x + x^2) + u_0 u_1 (2 + x + 2 x^2)(1 + 2 x + 3 x^2) +  u_1^2 x^2 (4 + 5 x + 13 x^2 + 2 x^3)) + \\
&+& u_1 (u_0 (1 + 2 x^2) + 3 u_1 x^2)(u_0 (1 + 2 x) + u_1 x^2 (2 + x)))^2 \nonumber
\label{eq:fluxdp2simpB}.
\end{eqnarray}
\end{small}

where $p_0$ is the solution of $dJ/dp=0$. Because the expression in Equation (\ref{eq:fluxdp2simp}) is always positive for $u_0>0$, $u_1>0$ and $x>1$, any roots of $dJ/dp=0$ will be minima for $J$ as a function of $p$.

Finally, the following expression can be obtained for $dJ/du_0$,
\begin{equation}
\frac{dJ}{du_0} = \frac{A_{3}}{B_{3}}
\label{eq:fluxdpsimp2},
\end{equation}
where
\begin{small}
\begin{eqnarray}
A_{3} &=& u_1^2[
u_0^2 (p^2 (1+x) (1+x^2) (1+4 x^2 - x^3 + 2 x^4) + p u_1 x^2 (5 + 13 x + 12 x^2 + 25 x^3 + 7 x^4 + 8 x^5 + 2 x^6) + \nonumber \\
&+& 2 u_1^2 x^3 (1 + 2x + 10 x^2 _ 8 x^3 + 4 x^4 + 2 x^5)) + 2 u_0 x (p^3 (1+x)^2 (1+x^2)(2+x+3 x^2) + \nonumber \\
&+& p^2 u_1 x (1+x) (6 + 8 x + 29 x^2 + 14 x^3 + 25 x^4 + 2 x^5) + p u_1^2 x^2 (4 + 15 x + 43 x^2 + 44 x^3 + 63 x^4 + 11 x^5) + \\
&+& 3 u_1^3 x^4 (2+x)(1+x+4 x^2)) + p^4 x (1+x)^3 (1+x^2)(2+x+3 x^2) + p^3 u_1 x^2 (1+x)^2 (8 + 9 x + 34 x^2 + 15 x^3 + 28 x^4 + 2 x^5) + \nonumber \\
&+& p^2 u_1^2 x^3 (1+x) (8 + 26 x + 67 x^2 + 64 x^3 + 83 x^4 + 16 x^5) + p u_1^3 x^5 (24 + 47 x + 98 x^2 + 85 x^3 + 32 x^4 + 2 x^5) +  3 u_1^4 x^7 (2+x) (3 + 2 x + x^2)] \nonumber
\label{eq:fluxdpsimp2A},
\end{eqnarray}
\end{small}
and
\begin{small}
\begin{eqnarray}
B_{3} &=& 2[p^2 (1+x)(2 u_0 (1 + x + x^2) + u_1 x (2 + x + 3 x^2)) + p(2 u_0^2 (1 + x + x^2) + u_0 u_1 (2 + x + 2 x^2)(1 + 2 x + 3 x^2) +  u_1^2 x^2 (4 + 5 x + 13 x^2 + 2 x^3)) + \nonumber \\
&+& u_1 (u_0 (1 + 2 x^2) + 3 u_1 x^2)(u_0 (1 + 2 x) + u_1 x^2 (2 + x))]^2
\label{eq:fluxdpsimp2B}.
\end{eqnarray}
\end{small}


\end{widetext}

\nocite{*}






\end{document}